\documentclass[preprint,preprintnumbers,amsmath,amssymb,10pt]{revtex4}
\usepackage[colorlinks,linkcolor=red,citecolor=blue]{hyperref}
\usepackage{epsf,epsfig}
\usepackage{bm}
\usepackage{graphicx}
\usepackage{natbib}
\usepackage{color}
\usepackage{mathrsfs}
\usepackage{float}
\usepackage{multirow}
\usepackage{makecell}
\usepackage{booktabs}

\newcommand{\commentold}[1]{}
\DeclareMathSymbol{:}{\mathpunct}{operators}{"3A}
\begin{document}

\title{\textbf{New Barrow holographic dark energy: cosmological dynamics and cosmic chronometer analysis}}

\author{Omid Azarakhsh\footnote{shahram.azarakhsh@yahoo.com}, Tayeb Golanbari\footnote{t.golanbary@uok.ac.ir}, Behrooz Malekolkalami\footnote{b.malakolkalami@uok.ac.ir} and Kh. Saaidi\footnote{ksaaidi@uok.ac.ir}}
\affiliation{\small{Department of Physics, University of Kurdistan, P.O. Box 66177-15175, Sanandaj, Iran}}

\date{\today}

\begin{abstract}
We formulate new Barrow holographic dark energy (NBHDE) as a second-order expansion of the Barrow entropy correction near the standard holographic dark energy (HDE) limit, adopting the future event horizon as the infrared cutoff. Writing the Barrow factor in exponential form identifies $(\Delta/2)\ln S_{\rm BH}$, rather than $\Delta$ alone, as the expansion parameter. In a flat universe containing pressureless matter and NBHDE without a cosmological constant, we derive the background equations, establish consistency with the integral definition of the future event horizon, and recover HDE exactly at $\Delta=0$. We assess this approximation against the unexpanded Barrow holographic dark energy (BHDE) density, locally and through independently normalized background evolutions. Across $0.50\leq c\leq1.20$ and $0\leq\Delta\leq0.003$, the maximum fractional density difference remains below $1\%$ over $-0.99\leq z\leq5$, while the normalized Hubble rates differ by less than $0.074\%$ over $0\leq z\leq5$. Within this domain, the Barrow correction modestly changes the expansion rate and acceleration transition but more strongly affects the dark energy equation of state and phantom crossing. At $c=0.80$, $w_{d0}$ shifts from $-1.031$ in HDE to $-0.900$ at $\Delta=0.003$. We perform a conditional comparison with 36 spectroscopic cosmic chronometer (CC) measurements. At fixed $H_0=67.4\,{\rm km\,s^{-1}\,Mpc^{-1}}$ and $\Omega_{m0}=0.30$, the minimum lies on the HDE boundary at $\Delta=0$ and $c=0.8457$. Minimizing over $c$ at $\Delta=0.003$ gives $c_{\min}=0.6879$, with $\Delta\chi_{\rm CC}^2\simeq1.0\times10^{-3}$. NBHDE therefore provides a self-consistent, quantitatively controlled framework near the HDE limit, showing that Barrow corrections can alter dark energy behavior even when their effect on $H(z)$ is largely compensated by changing $c$.
\end{abstract}



\maketitle

\section{Introduction}
\label{secintro}

Observations of Type Ia supernovae established the accelerated expansion of the Universe and made dark energy a central problem in cosmology~\cite{Riess_1998,Perlmutter:1998np}. The evidence is supported by cosmic microwave background observations from the \emph{Planck} satellite~\cite{Planck2018}, baryon acoustic oscillation measurements~\cite{bao-key}, and probes of the expansion and growth histories, including cosmic chronometers, weak lensing, and large-scale structure~\cite{Copeland2006DEreview}. Although the cosmological constant gives the simplest successful description within $\Lambda$CDM, its vacuum energy scale and the coincidence problem motivate dynamical alternatives~\cite{Copeland2006DEreview,Bamba:2012cp}, including interacting or emergent dark sectors~\cite{DiValentino2020DarkSector,Yang2021EmergentDE} and nonstandard scalar field scenarios~\cite{Mohammadi2015,Aghamohammadi2013,10.1093/ptep/ptag057}.

Holographic dark energy (HDE) relates the cosmological energy density to the holographic principle, according to which the information content of a gravitating region is associated with its boundary~\cite{tHooft:1993dmi,Susskind:1994vu,Bousso:2002ju}. Combined with the relation between ultraviolet and infrared cutoffs proposed by Cohen \emph{et al.}~\cite{Cohen:1998zx}, this idea leads to a dark energy density determined by a cosmological infrared cutoff~\cite{Li:2004rb,Wang:2016och}. The future event horizon yields standard HDE and allows accelerated expansion at late times~\cite{Li:2004rb,Wang:2016och,MANOHARAN2025170231}. Interacting models, the coincidence problem, observational constraints, perturbation growth, generalized cutoffs, and scalar-tensor extensions have been studied extensively~\cite{PAVON2005206,Mehrabi:2015hva,Wang:2005ph,Huang:2004wt,Kao:2005CMB,Nojiri:2005pu,Saaidi2013}.

The thermodynamic basis of HDE makes generalized horizon entropies a natural source of modified models. Quantum and statistical corrections to the Bekenstein--Hawking area law have been considered in gravitational and cosmological settings~\cite{Das:2007mj,Radicella:2010ss,Nojiri:2021iko,ODINTSOV2023101159}. Relevant examples include Tsallis entropy~\cite{Tsallis:1987eu,PhysRevD.103.083505,Pandey:2021NTHDE}, Rényi entropy~\cite{Renyi:1961,Golanbari:2020coz,Tamri2026}, Kaniadakis statistics~\cite{Kaniadakis:2002zz,Kaniadakis:2005zk,Drepanou:2021jiv,Hernandez-Almada:2021aiw}, and Sharma--Mittal entropy~\cite{SharmaMittal:1975,SharmaMittal:1977}. These approaches examine how departures from the area law are transmitted to cosmological evolution through the holographic density.

Barrow entropy provides a geometrically motivated example. A deformation of the horizon surface gives $S_B\propto S_{\rm BH}^{1+\Delta/2}$, with the Bekenstein--Hawking limit recovered at $\Delta=0$~\cite{Barrow:2020tzx}. Its holographic implementation leads to the Barrow holographic dark energy (BHDE) dependence $\rho_d\propto L^{\Delta-2}$~\cite{Saridakis:2020zol}. Thermodynamic consistency, nonflat geometries, interacting and noninteracting models based on the future event horizon, and phase space behavior have since been examined~\cite{Saridakis:2020cqq,Adhikary:2021xym,Sudharani2024Barrow,Sudharani2026PhaseSpace}.

The cosmological behavior and inferred constraints of BHDE depend on the infrared cutoff, parameter domain, and data combination. Models using the future event horizon have been compared with supernova and cosmic chronometer (CC) data~\cite{Anagnostopoulos:2020ctz}, while the Granda--Oliveros cutoff has been studied at the background and observational levels~\cite{Nandhida:2021vxl,Oliveros:2022biu}. Massive neutrinos, broader data combinations, and interacting or nonflat extensions have also been considered~\cite{Mahmoudifard2024BarrowNeutrinosGO,Yarahmadi2024BarrowHubbleTensionGO,Adhikary2025InteractingBarrowNonflat}. Bayesian comparisons indicate that the outcome can differ when Barrow entropy modifies the holographic density or instead enters through horizon thermodynamics~\cite{Tyagi2024EntropicCosmologyBayesian}. Published results consequently range from consistency with the HDE boundary to regions favoring nonzero deformation~\cite{Anagnostopoulos:2020ctz,Sudharani2026PhaseSpace,Li2026HubbleTensionHDE,Oliveros:2022biu,Yarahmadi2024BarrowHubbleTensionGO}; recent studies using DESI DR2 data further illustrate this dependence on formulation and data~\cite{Luciano2026BarrowTsallisDESIDR2,Luciano2026BarrowTsallisGO}.

Most BHDE studies retain the full Barrow power law. Expansions about the HDE limit have previously been used to construct new Tsallis and new Rényi HDE models~\cite{Pandey:2021NTHDE,Tamri2026}, showing the value of isolating the leading corrections generated by generalized entropy relations. These results provide a clear motivation for the present aim: to develop and quantitatively test a controlled formulation for Barrow entropy near the HDE limit. The Barrow case nevertheless requires an independent derivation because its power law generates a logarithmic series specific to this entropy, and the accuracy of any finite truncation must be established separately. Logarithmic terms have also appeared in Barrow cosmology and models with generalized cutoffs~\cite{Luciano2025BHDEfGT,Bekova2025BHDE}. A recent logarithmically corrected BHDE model adds a separate logarithmic term to Barrow entropy and uses the Hubble horizon~\cite{KOTAL2026170657}. The present construction is distinct from these models: its logarithms arise solely from expanding the Barrow power law, rather than from adding a new entropy correction, and the infrared cutoff remains the future event horizon.

Writing the Barrow correction in exponential form identifies $(\Delta/2)\ln S_{\rm BH}$, rather than $\Delta$ alone, as the expansion parameter. This distinction matters because a cosmological horizon has a large entropy. We define new Barrow holographic dark energy (NBHDE) by retaining all terms through second-order in this quantity. Their coefficients are fixed by the Barrow expression and introduce no additional parameter. The model contains pressureless matter and NBHDE in a spatially flat universe, with no independent cosmological constant.

The contribution of this work is therefore a controlled formulation near the HDE limit, rather than a new entropy or cutoff. Starting from an explicit entropy ratio normalization, we derive the background equations, prove positivity of the density, recover HDE exactly at $\Delta=0$, and show that the present density constraint selects a unique positive horizon radius. We also establish the terminal condition required by the integral definition of the future event horizon for NBHDE when the second-order density is retained throughout the future evolution. The accuracy of the approximation is evaluated both locally, by comparing the two densities at the same horizon radius, and dynamically, by comparing independently normalized background evolutions over a finite redshift interval. The exact BHDE density is used only as a reference for this comparison; the terminal condition proved for NBHDE is not transferred to the reference evolution. This distinction separates the internal consistency of the model from the finite domain in which it approximates the full Barrow expression.

We then study the evolution of $\Omega_d$, $w_d$, and the deceleration parameter $q$, including the acceleration transition, crossing of the phantom divide, and joint dependence on $(c,\Delta)$. The expansion histories are finally compared conditionally with 36 spectroscopic CC  measurements at fixed $H_0$ and $\Omega_{m0}$, with minimization over $c$ at each fixed $\Delta$. This comparison is not a global parameter inference. It shows that a change produced by $\Delta$ in the expansion history can be largely compensated by changing $c$, even when the equation of state and the phantom crossing remain sensitive to the deformation.

The paper is organized as follows. Section~\ref{Model} defines NBHDE, derives its background equations, and specifies the exact BHDE reference density and the HDE limit. Section~\ref{cosmo_evol} presents the validity analysis and cosmological evolution. Section~\ref{sec:cc_analysis} describes the CC data and conditional comparison. Section~\ref{conclu} summarizes the conclusions, and Appendix~\ref{app:cc_data} lists the CC compilation.

\section{New Barrow holographic dark energy model}
\label{Model}

We define new Barrow holographic dark energy (NBHDE) by retaining the Barrow entropy correction through second-order in $(\Delta/2)\ln S_{\rm BH}$ about the standard holographic dark energy (HDE) limit. We then derive its background dynamics in a spatially flat universe containing pressureless matter and NBHDE. Radiation is neglected over the late time interval considered here. Throughout, we use natural units with the speed of light, $\hbar$, and $k_B$ set to unity, and adopt the reduced Planck mass convention $M_p^2=(8\pi G)^{-1}$.

\subsection{Barrow entropy and BHDE normalization}
\label{subsec:barrow_exact_parent}

For a spherical horizon of radius $L$ and area $A=4\pi L^2$, the Bekenstein--Hawking entropy is~\cite{Bekenstein1973,Hawking1975}
\begin{equation}\label{eq:bh_entropy}
	S_{\rm BH}=\frac{A}{4G}=8\pi^2M_p^2L^2.
\end{equation}
Barrow introduced the dimensionless parameter $\Delta$ to describe a deformation of the horizon geometry and its effect on the entropy~\cite{Barrow:2020tzx}. We choose the reference area in the Barrow relation to be $4G$. The resulting dimensionless entropy can be written as
\begin{equation}\label{eq:barrow_entropy_bh}
		S_B=\left(\frac{A}{4G}\right)^{1+\Delta/2}=S_{\rm BH}^{\,1+\Delta/2}=S_{\rm BH}\exp\!\left[\frac{\Delta}{2}\ln S_{\rm BH}\right], \qquad 0\leq\Delta\leq1.
\end{equation}
This normalization recovers the Bekenstein--Hawking entropy exactly at $\Delta=0$.

We adopt a holographic normalization in which the standard HDE density is multiplied by the entropy ratio $S_B/S_{\rm BH}$. The exact BHDE density associated with this prescription is
\begin{equation}\label{eq:exact_bhde_density}
	\rho_d^{\rm ex}(L)=\frac{3c^2M_p^2}{L^2}\frac{S_B}{S_{\rm BH}}=\frac{3c^2M_p^2}{L^2}\exp\!\left[\frac{\Delta}{2}\ln S_{\rm BH}\right],
\end{equation}
where $c>0$ is the dimensionless holographic parameter. Both $c$ and $\Delta$ are taken to be constant. Using Eq.~\eqref{eq:bh_entropy}, Eq.~\eqref{eq:exact_bhde_density} becomes
\begin{equation}\label{eq:exact_bhde_powerlaw}
	\rho_d^{\rm ex}(L)=3c^2M_p^2\left(8\pi^2M_p^2\right)^{\Delta/2}L^{\Delta-2}.
\end{equation}
This has the conventional BHDE dependence $\rho_d\propto L^{\Delta-2}$~\cite{Saridakis:2020zol}. The entropy ratio prescription is an explicit normalization choice rather than a unique consequence of the holographic bound. It fixes the amplitude relative to standard HDE without introducing an additional free parameter, and the same convention is used for every value of $\Delta$. At $\Delta=0$, Eq.~\eqref{eq:exact_bhde_density} reduces to $\rho_d^{\rm HDE}=3c^2M_p^2/L^2$ with the same value of $c$. The superscript ``ex'' denotes the exact density associated with this normalization. Its independently normalized background evolution will be used only as a finite interval reference for evaluating the second-order formulation.

\subsection{second-order formulation of NBHDE}
\label{subsec:nbhde_second_order}

At fixed $L$, expansion of Eq.~\eqref{eq:barrow_entropy_bh} gives
\begin{equation}\label{eq:barrow_entropy_second_order}
	\frac{S_B}{S_{\rm BH}}=1+\frac{\Delta}{2}\ln S_{\rm BH}+\frac{\Delta^2}{8}\ln^2 S_{\rm BH}+\mathcal O\!\left(\left[\frac{\Delta}{2}\ln S_{\rm BH}\right]^3\right).
\end{equation}
The expansion is controlled by $(\Delta/2)\ln S_{\rm BH}$ rather than by $\Delta$ alone. This distinction is important on cosmological scales, where the horizon entropy is large.

Retaining all terms through second-order, we define
\begin{equation}\label{eq:nbhde_correction_factor}
	\mathcal F_2(L,\Delta)\equiv1+\frac{\Delta}{2}\ln S_{\rm BH}+\frac{\Delta^2}{8}\ln^2 S_{\rm BH},
\end{equation}
and the NBHDE density is
\begin{equation}\label{eq:nbhde_density}
	\rho_d(L)=\frac{3c^2M_p^2}{L^2}\mathcal F_2(L,\Delta).
\end{equation}
Hereafter, $\rho_d$ without a superscript denotes the density in Eq.~\eqref{eq:nbhde_density}. The correction factor satisfies
\begin{equation}\label{eq:nbhde_density_positivity}
	\mathcal F_2=\frac{1}{2}\left\{\left[1+\frac{\Delta}{2}\ln S_{\rm BH}\right]^2+1\right\}\geq\frac{1}{2},
\end{equation}
so the NBHDE density is positive for every $L>0$. We solve the background equations using Eq.~\eqref{eq:nbhde_density} without further expansion of ratios or other nonlinear functions of $\mathcal F_2$. The resulting observables are therefore predictions of the model defined by the second-order density, rather than separately truncated Taylor series in $\Delta$.

At a common horizon radius and for the same parameter values, the local fractional difference from the exact BHDE density is
\begin{equation}\label{eq:nbhde_local_truncation_error}
	\delta_{\rm tr}(L)\equiv\left|\frac{\rho_d(L)}{\rho_d^{\rm ex}(L)}-1\right|=\left|\mathcal F_2(L,\Delta)\exp\!\left[-\frac{\Delta}{2}\ln S_{\rm BH}\right]-1\right|.
\end{equation}
This quantity measures a local density error at fixed $L$; it does not determine the difference between two independently evolved expansion histories. The latter comparison is defined and evaluated in Sec.~\ref{cosmo_evol}.

\subsection{Background dynamics and consistency}
\label{subsec:nbhde_dynamics}

We consider a spatially flat FLRW background with $H\equiv\dot a/a>0$ and normalize the scale factor by $a(t_0)=1$. The Friedmann equations for pressureless matter, $p_m=0$, and dark energy, $p_d=w_d\rho_d$, are
\begin{equation}\label{eq:friedmann_equations}
	\begin{aligned}
		3M_p^2H^2&=\rho_m+\rho_d,\\
		-2M_p^2\dot H&=\rho_m+\rho_d+p_d.
	\end{aligned}
\end{equation}
The density parameters obey
\begin{equation}\label{eq:density_parameters}
	\Omega_m\equiv\frac{\rho_m}{3M_p^2H^2}, \quad \Omega_d\equiv\frac{\rho_d}{3M_p^2H^2}, \quad \Omega_m+\Omega_d=1.
\end{equation}
Assuming no interaction between the two components,
\begin{equation}\label{eq:continuity_equations}
	\dot\rho_m+3H\rho_m=0, \qquad \dot\rho_d+3H(1+w_d)\rho_d=0,
\end{equation}
and hence $\rho_m=\rho_{m0}a^{-3}$.

Following the event horizon prescription of HDE~\cite{Li:2004rb}, the infrared cutoff is identified with the proper radius of the future event horizon:
\begin{equation}\label{eq:future_event_horizon}
	\begin{aligned}
		L(t)&=a(t)\int_t^{t_f}\frac{dt'}{a(t')},\\
		\dot L&=HL-1.
	\end{aligned}
\end{equation}
Here $t_f=\infty$ when the expansion extends to infinite cosmic time; if the scale factor diverges at a finite time endpoint, that endpoint is used instead. In either case, the integral must converge. The second relation in Eq.~\eqref{eq:future_event_horizon} follows by differentiation, while its converse also requires the terminal condition discussed below.

Combining Eq.~\eqref{eq:nbhde_density} with the Friedmann constraint gives
\begin{equation}\label{eq:omega_d_constraint}
	\Omega_d=\frac{c^2\mathcal F_2(L,\Delta)}{H^2L^2}, \qquad \frac{1}{HL}=\frac{\sqrt{\Omega_d}}{c\sqrt{\mathcal F_2(L,\Delta)}}.
\end{equation}
Since $d\ln S_{\rm BH}/d\ln L=2$, differentiation of the NBHDE density, together with the dark energy conservation equation and $\dot L=HL-1$, yields
\begin{equation}\label{eq:nbhde_eos}
	w_d=-1+\frac{1}{3}\left[2-\frac{\Delta+\frac{\Delta^2}{2}\ln S_{\rm BH}}{\mathcal F_2(L,\Delta)}\right]\left(1-\frac{1}{HL}\right).
\end{equation}
Thus, the pressure is determined by the density and the horizon evolution; no independent equation of state is imposed. Applying the same conservation prescription to the independently evolved exact density reference background gives
\begin{equation}\label{eq:exact_bhde_eos}
	w_d^{\rm ex}=-1+\frac{2-\Delta}{3}\left(1-\frac{1}{H^{\rm ex}L^{\rm ex}}\right),
\end{equation}
where $H^{\rm ex}$ and $L^{\rm ex}$ denote the corresponding reference background quantities.

For $L>0$ and $0\leq\Delta\leq1$, the coefficient of $1-1/(HL)$ in Eq.~\eqref{eq:nbhde_eos} has a fixed positive sign. Explicitly,
\begin{equation}\label{eq:phantom_condition}
	\begin{aligned}
		2-\frac{\Delta+\frac{\Delta^2}{2}\ln S_{\rm BH}}{\mathcal F_2(L,\Delta)}
		&=\frac{\left[\Delta\ln S_{\rm BH}+(2-\Delta)\right]^2+4-\Delta^2}{4\mathcal F_2(L,\Delta)}>0,\\
		HL<1&\Longleftrightarrow w_d<-1,\\
		HL=1&\Longleftrightarrow w_d=-1,\\
		HL>1&\Longleftrightarrow w_d>-1.
	\end{aligned}
\end{equation}
A phantom divide crossing therefore occurs when $HL$ passes through unity. Since $2-\Delta>0$ over the same range, Eq.~\eqref{eq:exact_bhde_eos} has the corresponding algebraic criterion in terms of $H^{\rm ex}L^{\rm ex}$. This local statement does not establish the future event horizon terminal condition for either background.

Writing $x\equiv\ln a=-\ln(1+z)$ and using a prime for $d/dx$, the closed autonomous system for $(\Omega_d,L)$ is
\begin{align}
	\Omega_d'&=-3w_d\Omega_d(1-\Omega_d),
	\label{eq:omega_d_autonomous}\\
	L'&=L\left[1-\frac{\sqrt{\Omega_d}}{c\sqrt{\mathcal F_2(L,\Delta)}}\right].
	\label{eq:L_autonomous}
\end{align}
The Hubble and deceleration parameters follow from
\begin{align}
	H&=\frac{c\sqrt{\mathcal F_2(L,\Delta)}}{L\sqrt{\Omega_d}},
	\label{eq:hubble_from_omega_L}\\
	q&\equiv-1-\frac{\dot H}{H^2}=\frac{1}{2}\left(1+3w_d\Omega_d\right).
	\label{eq:deceleration_parameter}
\end{align}

\subsubsection{Present normalization}
\label{subsec:present_horizon_scale}

At the present epoch, we specify $H_0>0$ and $0<\Omega_{d0}=1-\Omega_{m0}<1$. For each choice of $(c,\Delta)$, the horizon radius $L_0\equiv L(a=1)$ is fixed by
\begin{equation}\label{eq:present_root}
	\frac{c^2\mathcal F_2(L_0,\Delta)}{H_0^2L_0^2}=\Omega_{d0}.
\end{equation}
The entropy entering $\mathcal F_2$ is evaluated using Eq.~\eqref{eq:bh_entropy}, with all dimensional quantities expressed in consistent natural units before $\ln S_{\rm BH}$ is calculated.

The positive solution of Eq.~\eqref{eq:present_root} is unique. Indeed,
\begin{equation}\label{eq:present_root_monotonicity}
	\frac{d}{d\ln L}\left[\frac{\mathcal F_2(L,\Delta)}{L^2}\right]
	=-\frac{\left[\Delta\ln S_{\rm BH}+(2-\Delta)\right]^2+4-\Delta^2}{4L^2}<0.
\end{equation}
For $0\leq\Delta\leq1$, the function $\mathcal F_2(L,\Delta)/L^2$ is therefore continuous and strictly decreasing on $L>0$. It diverges as $L\to0^+$ and tends to zero as $L\to\infty$, establishing the existence and uniqueness of a positive $L_0$.

For the exact density reference evolution, the same present values of $(H_0,\Omega_{d0},c,\Delta)$ are imposed independently. Equation~\eqref{eq:exact_bhde_powerlaw} then gives
\begin{equation}\label{eq:exact_present_root}
	L_0^{\rm ex}=\left[\frac{c^2\left(8\pi^2M_p^2\right)^{\Delta/2}}{H_0^2\Omega_{d0}}\right]^{1/(2-\Delta)}.
\end{equation}
Because $2-\Delta>0$, this is also the unique positive root of its density constraint. At $\Delta=0$, both present radii reduce to $c/(H_0\sqrt{\Omega_{d0}})$.

The present phantom boundary follows from $H_0L_0=1$ and Eq.~\eqref{eq:present_root}:
\begin{equation}\label{eq:c_phantom_divide}
	c=\left[\frac{\Omega_{d0}}{\mathcal F_2(H_0^{-1},\Delta)}\right]^{1/2}, \qquad w_{d0}=-1.
\end{equation}
At fixed $\Delta$, values of $c$ below this boundary give $w_{d0}<-1$, whereas values above it give $w_{d0}>-1$. The uniqueness statements above concern only the algebraic normalization at the present epoch; in particular, they do not establish the terminal condition for the exact density reference evolution.

\subsubsection{Consistency of the horizon prescription}
\label{subsec:horizon_terminal_condition}

The local relation $\dot L=HL-1$ implies $d(L/a)/dt=-1/a$. It reproduces the integral definition in Eq.~\eqref{eq:future_event_horizon} only if
\begin{equation}\label{eq:horizon_terminal_condition}
	\lim_{t\to t_f}\frac{L(t)}{a(t)}=0.
\end{equation}
Thus, the local horizon equation is necessary but not sufficient by itself; the terminal condition removes its integration constant.

For NBHDE, the condition can be established if the second-order density in Eq.~\eqref{eq:nbhde_density} is retained throughout the future expansion. Nonnegative matter density and Eq.~\eqref{eq:nbhde_density_positivity} give
\begin{equation}\label{eq:horizon_growth_bounds}
	\begin{aligned}
		(HL)^2&=\frac{\rho_mL^2}{3M_p^2}+c^2\mathcal F_2(L,\Delta)\geq\frac{c^2}{2},\\
		1-\frac{\sqrt{2}}{c}&\leq\frac{d\ln L}{d\ln a}\leq1.
	\end{aligned}
\end{equation}
For $a\geq1$, these bounds keep $L$ positive and finite at every finite $a$, imply $L\leq L_0a$, and restrict $|\ln(L/L_0)|$ to grow at most linearly with $\ln a$. Integrating the horizon equation gives
\begin{equation}\label{eq:comoving_horizon_solution}
	\frac{L(a)}{a}=L_0\exp\!\left[-\int_1^a\frac{da'}{a'H(a')L(a')}\right].
\end{equation}
Since $\rho_m=\rho_{m0}a^{-3}$ and $L\leq L_0a$, the matter contribution to $(HL)^2$ is bounded by $\rho_{m0}L_0^2/(3M_p^2a)$. Moreover, $\mathcal F_2$ grows at most quadratically with $\ln a$. Hence there is a finite positive constant $K$, independent of $a$, such that $HL\leq K(1+\ln a)$, and
\begin{equation}\label{eq:horizon_integral_bound}
	\int_1^a\frac{da'}{a'H(a')L(a')}\geq\frac{\ln(1+\ln a)}{K}\longrightarrow\infty\quad(a\to\infty).
\end{equation}
Equation~\eqref{eq:comoving_horizon_solution} then gives $L/a\to0$, establishing Eq.~\eqref{eq:horizon_terminal_condition}.

This proof concerns the NBHDE model defined by the second-order density and assumes that the same polynomial density is retained throughout the future evolution. It establishes the internal consistency of the NBHDE horizon prescription, but not the accuracy of the entropy expansion at arbitrarily late times. It also does not establish the terminal condition for the independently normalized exact density reference evolution, which is used here only over the stated finite redshift interval.

\subsubsection{HDE limit}
\label{subsec:nbhde_hde_limit}

At $\Delta=0$, $\mathcal F_2(L,0)=1$, and both the NBHDE and exact reference densities reduce to standard HDE. The background relations become
\begin{equation}\label{eq:hde_limit_relations}
	\begin{aligned}
		\frac{1}{HL}&=\frac{\sqrt{\Omega_d}}{c},\\
		w_d^{\rm HDE}&=-\frac{1}{3}-\frac{2\sqrt{\Omega_d}}{3c},\\
		\Omega_d'&=\Omega_d(1-\Omega_d)\left(1+\frac{2\sqrt{\Omega_d}}{c}\right).
	\end{aligned}
\end{equation}
The present horizon radius simultaneously reduces to $L_0=c/(H_0\sqrt{\Omega_{d0}})$. These are the standard future event horizon HDE relations~\cite{Li:2004rb}, and their exact recovery provides analytic checks of both the formulation and its numerical implementation.

\section{Cosmological evolution and numerical results}
\label{cosmo_evol}

Using the background system derived in Sec.~\ref{Model}, we investigate the evolution of the dark energy fraction, the equation of state, and the cosmic acceleration as the parameters $c$ and $\Delta$ are varied. Two distinct comparisons are used throughout this section. The exact BHDE density provides a reference for quantifying the accuracy of the second-order formulation, whereas standard HDE identifies the physical changes produced by a nonzero Barrow deformation. Keeping these roles separate is essential: agreement with the exact density reference measures truncation accuracy, while departure from HDE describes the dynamics of NBHDE itself.

\subsection{Accuracy and domain of the second-order formulation}
\label{subsec:numerical_setup}

For the numerical analysis, we adopt the fixed background normalization
\begin{align}\label{eq:present_normalization}
	H_0&=67.4\,{\rm km\,s^{-1}\,Mpc^{-1}},\nonumber\\
	\Omega_{m0}&=0.30,\qquad \Omega_{d0}=0.70.
\end{align}
These quantities are held fixed rather than inferred from the data. For every pair $(c,\Delta)$, the NBHDE horizon radius is initialized with the unique positive solution of Eq.~\eqref{eq:present_root}. The reference evolution based on the exact density is normalized independently using Eq.~\eqref{eq:exact_present_root}, with the same values of $(H_0,\Omega_{m0},\Omega_{d0},c,\Delta)$.

The numerical scan covers $0.50\leq c\leq1.20$ with a spacing of $0.01$, together with $\Delta=0$, $0.001$, $0.002$, $0.003$, $0.004$, and $0.005$. The background solutions are evaluated over $-0.99\leq z\leq5$, while the evolution curves are displayed over $-0.9\leq z\leq3$. The cases $\Delta=0.004$ and $0.005$ are included only to show how the second-order description deteriorates beyond the range adopted for the main analysis. Since radiation is not included, the calculations are restricted to the late time interval ending at $z=5$ and are not extrapolated to the early Universe.

We assess the accuracy of the formulation at two complementary levels. The local density error $\delta_{\rm tr}$, defined in Eq.~\eqref{eq:nbhde_local_truncation_error}, compares the second-order and exact BHDE densities at the same horizon radius. It is evaluated along the NBHDE horizon solution and therefore tests the local truncation of the density relation. To compare the independently evolved expansion histories, we use
\begin{equation}\label{eq:num_variables}
	E(z)\equiv\frac{H(z)}{H_0},
\end{equation}
and define
\begin{equation}\label{eq:perturbative_diagnostics}
	\delta_E(z)\equiv\left|\frac{E_{(2)}(z)}{E_{\rm ex}(z)}-1\right|.
\end{equation}
Here, the subscripts $(2)$ and ${\rm ex}$ denote the independently normalized NBHDE and exact density reference evolutions, respectively. Unlike $\delta_{\rm tr}$, the quantity $\delta_E$ includes the cumulative response of the background evolution to the two density prescriptions. Neither diagnostic measures the physical departure of NBHDE from standard HDE.

\begin{figure*}
	\centering
	\includegraphics[width=0.92\textwidth]{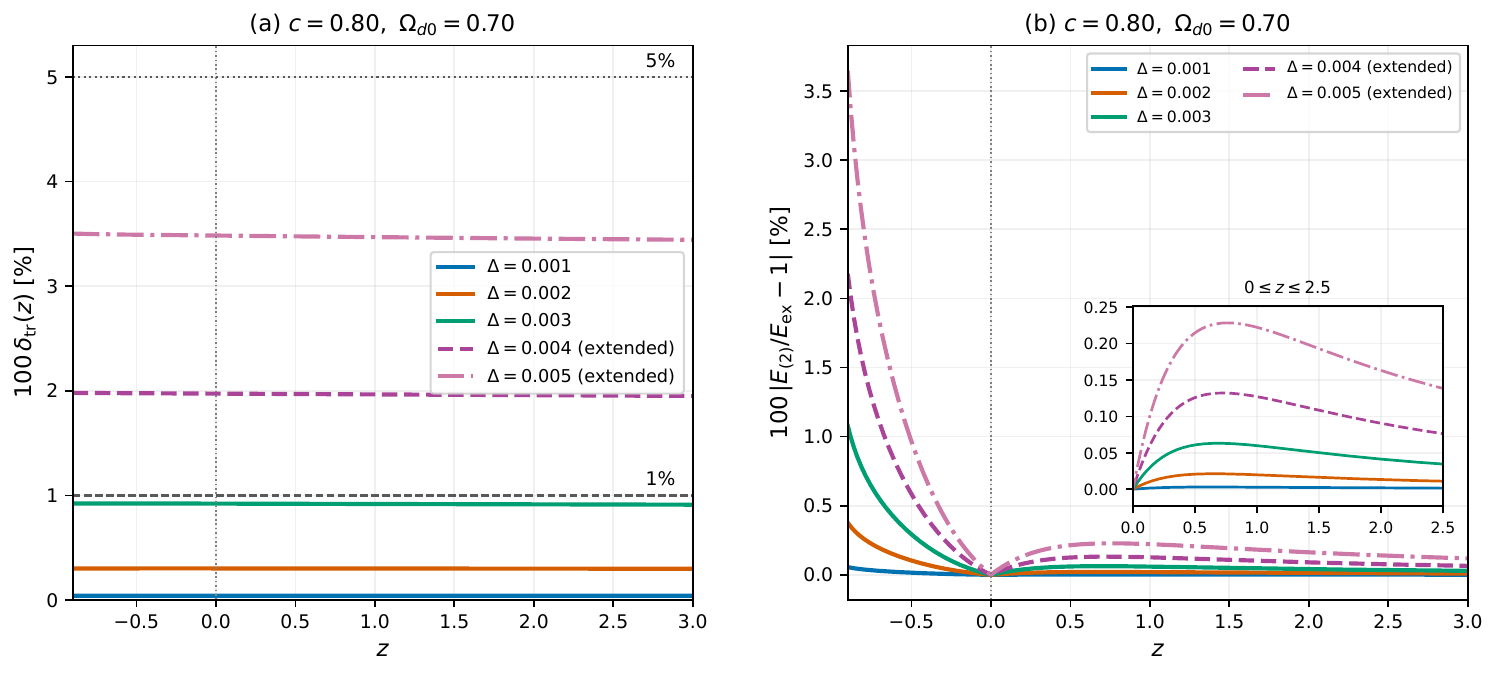}
	\caption{Accuracy of the second-order formulation relative to the exact BHDE density reference for $c=0.80$ and $\Omega_{d0}=0.70$. Panel (a) shows the local density error $100\delta_{\rm tr}$ evaluated at the NBHDE horizon radius. Panel (b) shows the expansion rate difference $100\delta_E$ between the independently normalized background evolutions. The curves are displayed over $-0.9\leq z\leq3$, and the inset enlarges the interval $0\leq z\leq2.5$. The dashed and dash dotted curves correspond to the extended cases $\Delta=0.004$ and $0.005$.}
	\label{fig:perturbative_validation}
\end{figure*}

The two measures are contrasted in Fig.~\ref{fig:perturbative_validation} for $c=0.80$. The local density error depends strongly on $\Delta$ but only weakly on redshift over the displayed interval. Its maximum values over $-0.99\leq z\leq5$ are approximately $0.042\%$, $0.303\%$, and $0.923\%$ for $\Delta=0.001$, $0.002$, and $0.003$, respectively. The extended cases $\Delta=0.004$ and $0.005$ lie clearly above the one percent level.
The expansion history over the past interval is reproduced more accurately than the local density error alone would suggest. For $c=0.80$, the maximum values of $\delta_E$ over $0\leq z\leq5$ are $0.0030\%$, $0.0212\%$, and $0.0631\%$ for $\Delta=0.001$, $0.002$, and $0.003$, respectively. The equality $E_{(2)}(0)=E_{\rm ex}(0)=1$ forces $\delta_E(0)=0$ and is therefore a normalization condition rather than an independent accuracy test. Toward higher positive redshift, the increasing matter contribution reduces the influence of the residual difference in the dark energy density on the total expansion rate.

Repeating the expansion rate comparison over the complete sampled grid with $0.50\leq c\leq1.20$ and $0\leq\Delta\leq0.003$ gives $100\delta_E^{\max}=0.0732\%$ over $0\leq z\leq5$. This largest value occurs at $c=0.50$ and $\Delta=0.003$, near $z=0.469$. Thus, throughout the sampled parameter grid, the past expansion rate difference remains below $0.074\%$.

This bound must not be extended to negative redshift. On the negative redshift side of the finite numerical interval, $\delta_E$ can grow appreciably even when the local density error remains below one percent. For example, at $c=0.80$ and $\Delta=0.003$, it reaches approximately $2.18\%$ by $z=-0.99$. Moreover, the terminal condition for the exact density reference evolution has not been established. Its negative redshift continuation is therefore used only as a finite interval diagnostic and is not assigned an independently established physical future interpretation.

\begin{figure}
	\centering
	\includegraphics[width=0.5\columnwidth]{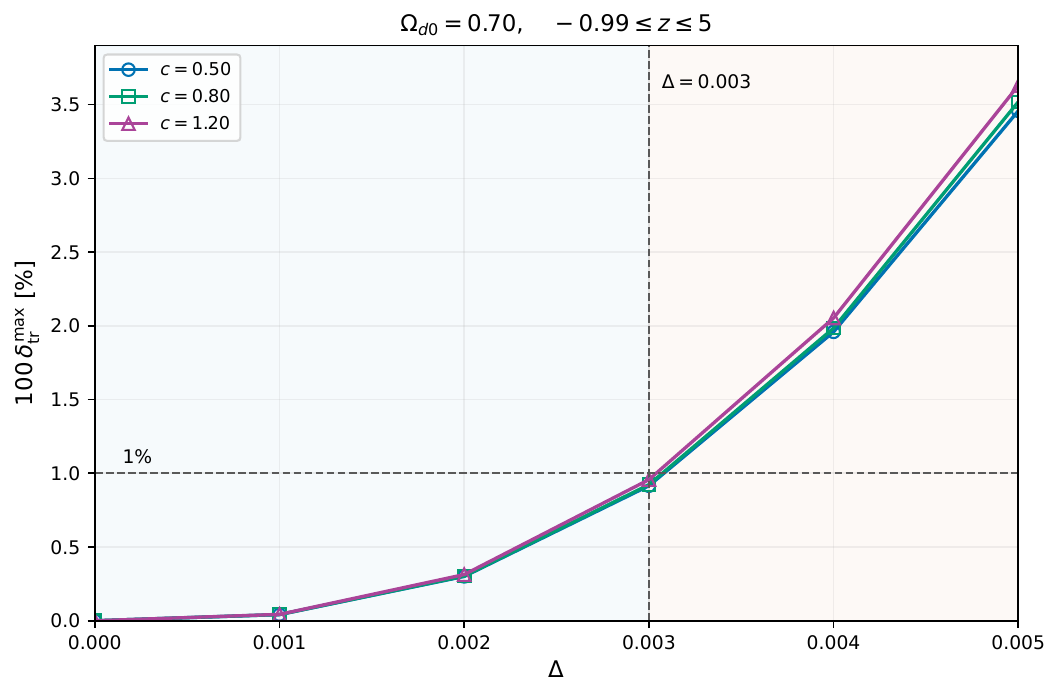}
	\caption{Maximum local density error $100\delta_{\rm tr}^{\max}$ over $-0.99\leq z\leq5$ as a function of $\Delta$, shown for $c=0.50$, $0.80$, and $1.20$. The horizontal dashed line marks the one percent criterion. The vertical dashed line at $\Delta=0.003$ marks the upper limit adopted for the principal analysis; values to its right are displayed only to illustrate the loss of accuracy beyond this range.}
	\label{fig:perturbative_domain}
\end{figure}

The dependence of the local density error on both model parameters is summarized in Fig.~\ref{fig:perturbative_domain}. Its variation with $c$ is weak but nonzero compared with its rapid growth with $\Delta$. Across the sampled $c$ grid, the largest value of $100\delta_{\rm tr}$ at $\Delta=0.003$ is $0.957\%$. By contrast, the ranges at $\Delta=0.004$ and $0.005$ are approximately $1.96$--$2.05\%$ and $3.46$--$3.63\%$, respectively.
We therefore adopt $0\leq\Delta\leq0.003$ for the principal cosmological analysis. This is a conservative working range defined by the subpercent local density error for the fixed background normalization, sampled values of $c$, and redshift interval considered here. It is not an observational constraint on $\Delta$ and should not be interpreted as a universal accuracy domain for other cosmologies, cutoff prescriptions, or derived observables.

Within this range, the past expansion history satisfies the stronger bound $100\delta_E^{\max}<0.074\%$ over $0\leq z\leq5$. Nevertheless, small differences in $E(z)$ do not guarantee agreement in quantities that depend sensitively on the condition $HL=1$, particularly the present phantom classification and the crossing redshift. These cases are examined explicitly below. The subsequent negative redshift curves describe the finite interval evolution of the NBHDE model itself; they are not used to claim that the second-order density remains an accurate representation of exact BHDE at arbitrarily late times.

\subsection{Evolution of the dark energy density parameter}
\label{subsec:omega_evolution}

\begin{figure*}
	\centering
	\includegraphics[width=0.92\textwidth]{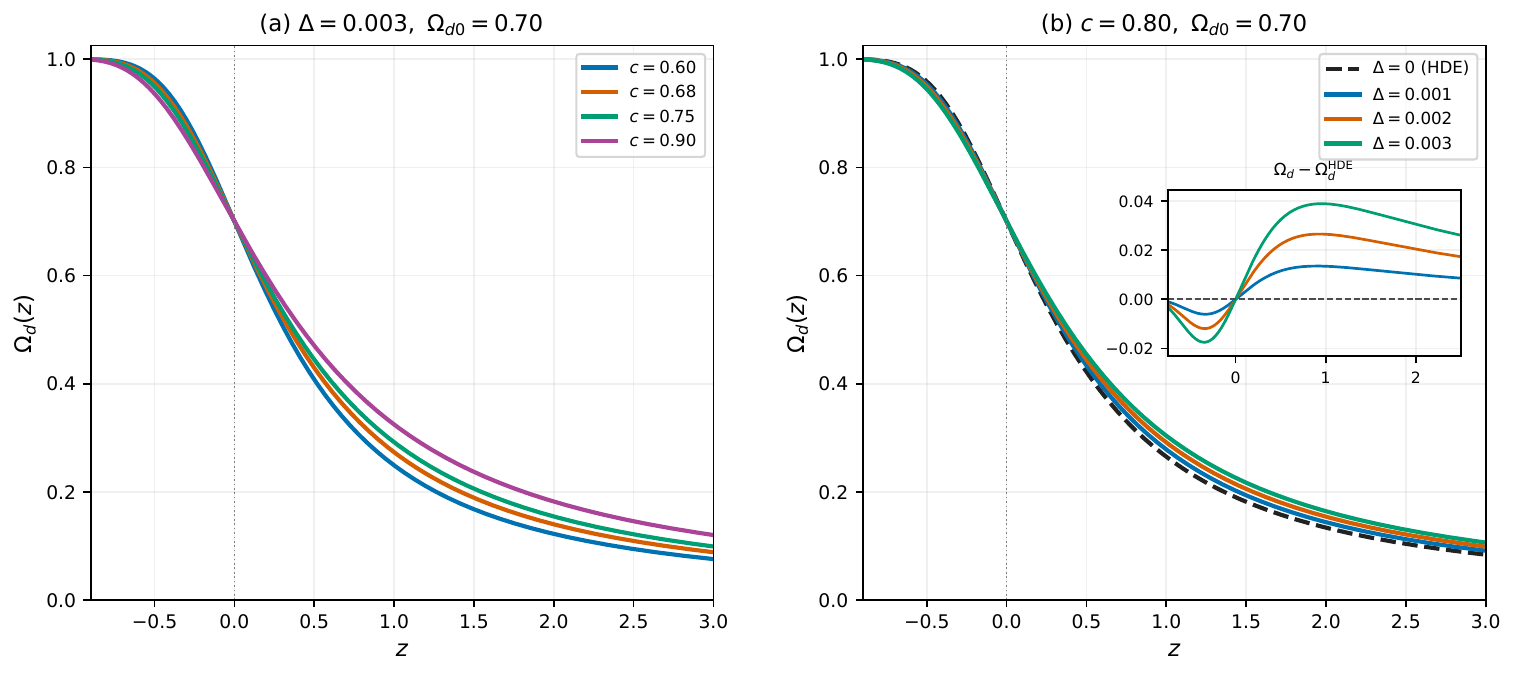}
	\caption{Evolution of $\Omega_d(z)$ for $\Omega_{d0}=0.70$. Panel (a) shows the dependence on $c$ at fixed $\Delta=0.003$, while panel (b) shows the dependence on $\Delta$ at fixed $c=0.80$. The dashed black curve represents the HDE limit. The inset displays the difference $\Omega_d-\Omega_d^{\rm HDE}$, with colors matching those in the main panel. The vertical dotted line marks the present epoch.}
	\label{fig:omega_evolution}
\end{figure*}

Figure~\ref{fig:omega_evolution} presents the evolution of the dark energy density parameter for the normalization in Eq.~\eqref{eq:present_normalization}. In panel (a), the holographic parameter is varied over $c=0.60$, $0.68$, $0.75$, and $0.90$ at fixed $\Delta=0.003$. Panel (b) instead fixes $c=0.80$ and compares $\Delta=0.001$, $0.002$, and $0.003$ with the HDE limit $\Delta=0$.
All curves satisfy $\Omega_d(0)=0.70$ by construction. Toward positive redshift, $\Omega_d$ decreases and the relative contribution of matter grows. Over the displayed negative redshift interval, the dark energy fraction increases and approaches unity. This behavior follows directly from the evolution equation \eqref{eq:omega_d_autonomous} for the negative values of $w_d$ realized by the solutions considered here.

At fixed $\Delta=0.003$, increasing $c$ produces a larger value of $\Omega_d$ at a given positive redshift. The ordering is reversed for $z<0$, where the models with larger $c$ evolve more slowly toward dark energy domination. Since every curve is constrained to pass through the same present value, their intersection at $z=0$ reflects the imposed normalization and does not imply an absence of dependence on $c$.

At fixed $c=0.80$, increasing $\Delta$ raises $\Omega_d$ relative to HDE for $z>0$ and lowers it for $z<0$. The differences are small in the main panel but are resolved clearly by the inset. The residual vanishes at the present epoch, reaches a broad positive maximum near $z\simeq1$, and decreases again toward higher redshift. On the negative redshift side, it becomes negative before its magnitude decreases near the lower end of the displayed interval, where all curves lie close to $\Omega_d=1$.

The residual shown in the inset measures the physical departure of NBHDE from the HDE limit and should not be confused with either of the approximation diagnostics introduced in Sec.~\ref{subsec:numerical_setup}. Although variations in $c$ and $\Delta$ can generate similar shifts in $\Omega_d(z)$ over part of the redshift range, their effects on the equation of state and the transition epochs need not be equivalent. We therefore examine the evolution of $w_d$ separately in the following subsection.

\subsection{Equation of state and phantom crossing}
\label{subsec:eos_phantom}

\begin{figure*}
	\centering
	\includegraphics[width=0.92\textwidth]{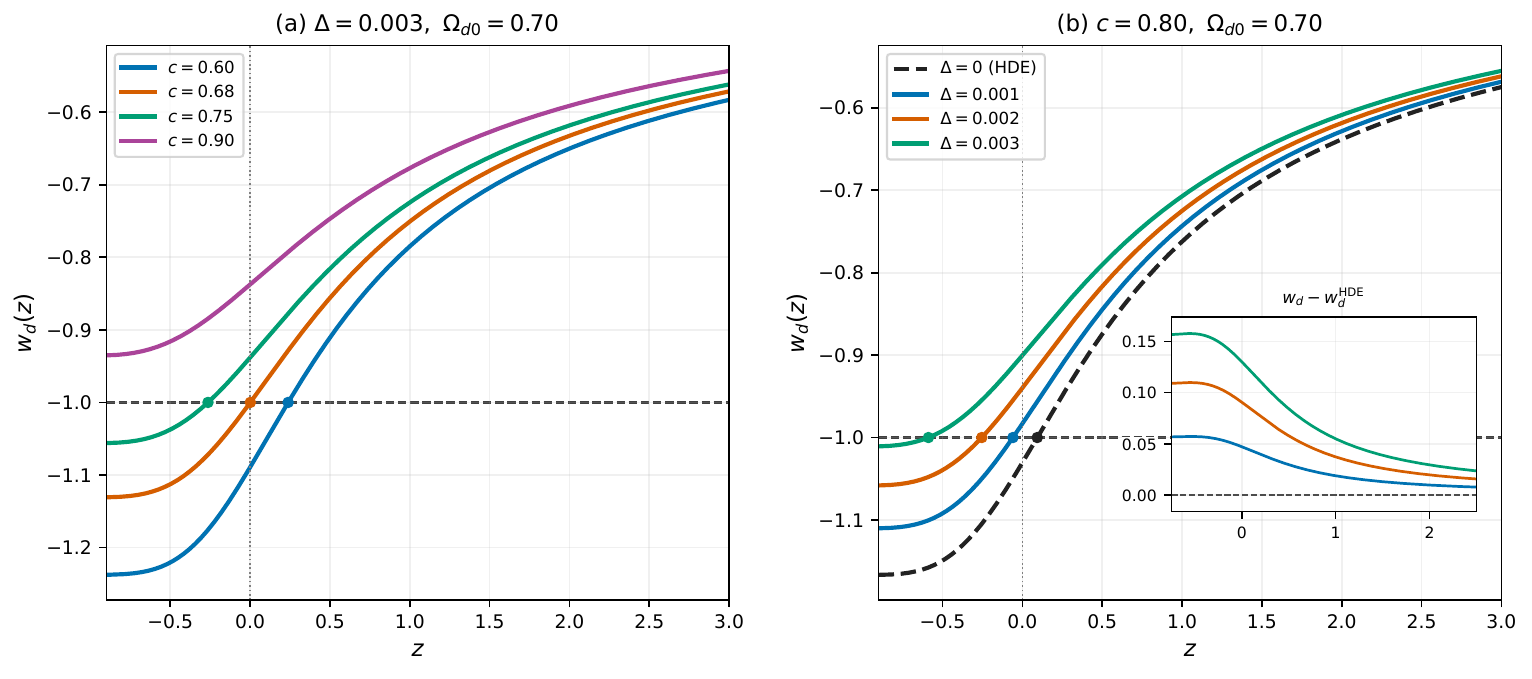}
	\caption{Evolution of the dark energy equation of state for $\Omega_{d0}=0.70$. Panel (a) shows the dependence on $c$ at fixed $\Delta=0.003$, while panel (b) shows the dependence on $\Delta$ at fixed $c=0.80$. The horizontal dashed line marks the phantom divide $w_d=-1$, and the filled circles identify its crossings. In panel (b), the dashed black curve represents the HDE limit and the inset displays $w_d-w_d^{\rm HDE}$. The vertical dotted line marks the present epoch.}
	\label{fig:w_evolution}
\end{figure*}

The dark energy equation of state is more sensitive to the model parameters than the density fraction discussed above. As established by Eq.~\eqref{eq:phantom_condition}, the sign of $w_d+1$ is determined by the value of $HL-1$: the dark energy is phantom for $HL<1$, nonphantom for $HL>1$, and reaches the phantom divide when $HL=1$. Figure~\ref{fig:w_evolution} shows how this behavior changes when $c$ and $\Delta$ are varied separately.
At fixed $\Delta=0.003$, increasing $c$ shifts the low redshift equation of state toward less negative values. For $c=0.60$, the present value is $w_{d0}=-1.0894$, and the phantom divide is crossed in the past at $z_{\rm ph}\simeq0.238$. The case $c=0.68$ lies very close to the divide, with $w_{d0}=-1.0004$ and $z_{\rm ph}\simeq0.001$. For $c=0.75$, the present dark energy is nonphantom, with $w_{d0}=-0.9382$, and the crossing occurs within the displayed future interval at $z_{\rm ph}\simeq-0.264$. Increasing the parameter further to $c=0.90$ gives $w_{d0}=-0.8374$, with no crossing found over $-0.99\leq z\leq5$.

The sequence in panel (a) shows that changing $c$ can move the crossing continuously from the past to the future and eventually outside the numerical interval. Whenever a crossing occurs for these representative models, the evolution proceeds from $w_d>-1$ to $w_d<-1$ as cosmic time increases. The absence of a crossing within the explored interval, however, does not determine the behavior at arbitrarily late times.

Panel (b) isolates the effect of the Barrow deformation at fixed $c=0.80$. In the HDE limit, $w_{d0}=-1.0306$, so the dark energy is mildly phantom at the present epoch. The values $\Delta=0.001$, $0.002$, and $0.003$ instead give $w_{d0}=-0.9833$, $-0.9399$, and $-0.9004$, respectively. Thus, even the smallest nonzero deformation shown in the figure moves the present solution to the nonphantom side of the divide. Over the same sequence, the crossing redshift changes from $z_{\rm ph}\simeq0.093$ in HDE to $-0.059$, $-0.254$, and $-0.588$. Within the adopted parameter range, the Barrow correction can therefore change both the present phantom classification and the temporal location of the crossing.

The inset confirms that increasing $\Delta$ produces a positive shift in $w_d$ relative to HDE throughout the displayed interval. This shift is most pronounced at low and negative redshift and becomes smaller toward the matter dominated regime. Unlike the residual in $\Omega_d$, it does not vanish at $z=0$, because the common normalization fixes the present density fraction but not the dark energy pressure. The inset represents the physical difference between NBHDE and HDE at fixed $c$ and should not be interpreted as a measure of the second-order truncation error.

Near the phantom divide, even a small difference between the second-order and exact density reference evolutions can affect the classification. For $c=0.68$ and $\Delta=0.003$, their independently normalized present values are $w_{d0}^{(2)}\simeq-1.000429$ and $w_{d0}^{\rm ex}\simeq-0.997355$. Although their absolute difference is small, the two values lie on opposite sides of $w_d=-1$. The NBHDE evolution crosses the divide at $z_{\rm ph}^{(2)}\simeq0.00143$, whereas the exact density reference evolution has a zero of $w_d+1$ at $z_{\rm ph}^{\rm ex}\simeq-0.00890$. This occurs even though the maximum relative difference between their normalized expansion rates is only about $0.0668\%$ over $0\leq z\leq5$.

A similar sensitivity appears in the crossing epoch even when the present classifications agree. For $c=0.80$ and $\Delta=0.003$, both evolutions are nonphantom today, but their zeros of $w_d+1$ occur at $z_{\rm ph}^{(2)}\simeq-0.5881$ and $z_{\rm ph}^{\rm ex}\simeq-0.6337$. The negative redshift results obtained from the exact density reference evolution are used only as finite interval diagnostics; they are not interpreted as independently established physical future crossings. The curves in Fig.~\ref{fig:w_evolution} describe the NBHDE model itself, while the reference comparison shows that close agreement in $H(z)$ does not necessarily guarantee the same phantom classification or crossing redshift.

\subsection{Cosmic acceleration and transition redshift}
\label{subsec:q_transition}

\begin{figure*}
	\centering
	\includegraphics[width=0.92\textwidth]{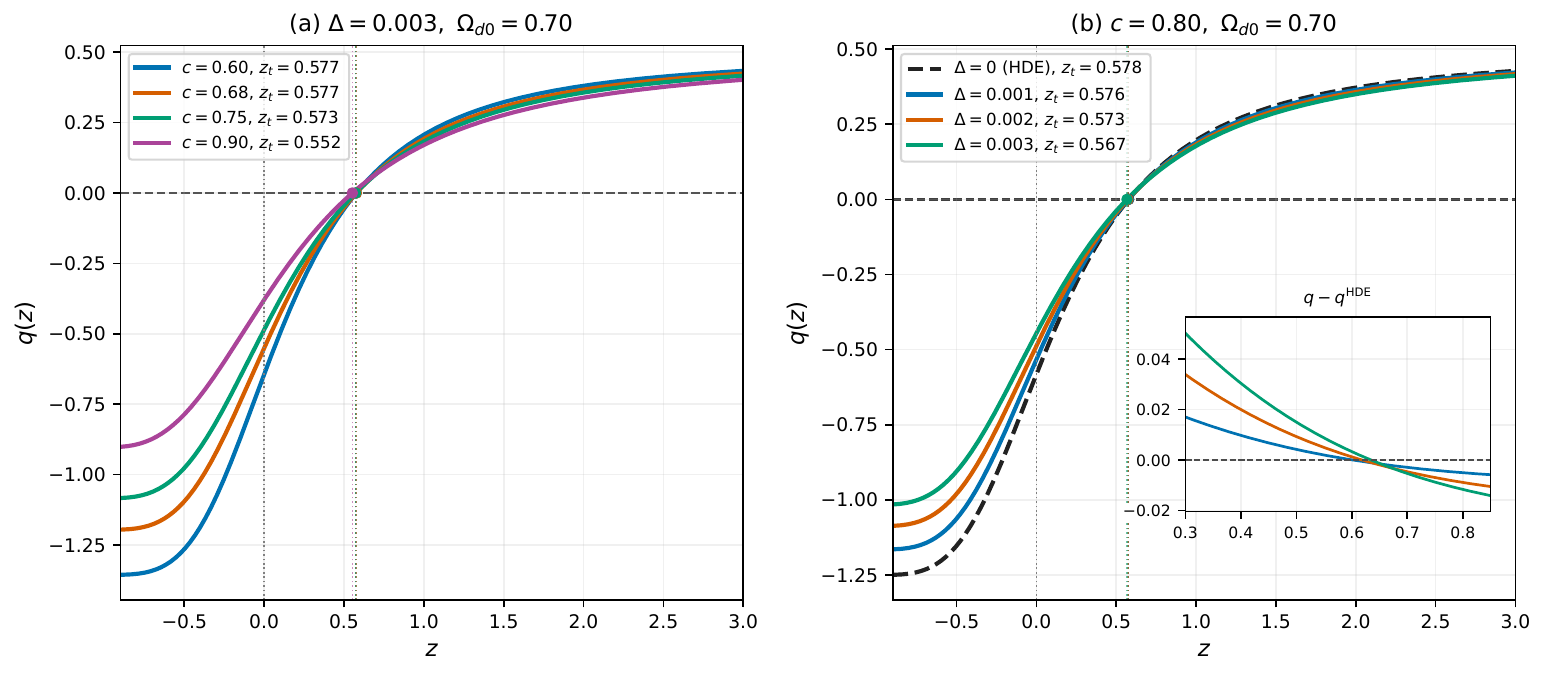}
	\caption{Evolution of the deceleration parameter for $\Omega_{d0}=0.70$. Panel (a) shows the dependence on $c$ at fixed $\Delta=0.003$, while panel (b) shows the dependence on $\Delta$ at fixed $c=0.80$. The horizontal dashed line marks $q=0$, and the filled circles and corresponding vertical dotted lines identify the transition redshifts listed in the legends. The gray vertical dotted line marks the present epoch. In panel (b), the dashed black curve represents the HDE limit and the inset displays $q-q^{\rm HDE}$ near the transition.}
	\label{fig:q_evolution}
\end{figure*}

We next examine the deceleration parameter obtained from Eq.~\eqref{eq:deceleration_parameter}. The transition redshift $z_t$ is defined by $q(z_t)=0$, with the expansion changing from deceleration to acceleration as cosmic time increases. Figure~\ref{fig:q_evolution} shows this evolution for the same representative parameter choices used in the preceding subsections.
Every displayed trajectory undergoes a single transition from decelerated expansion at positive redshift to accelerated expansion before the present epoch. Toward the matter dominated part of the displayed range, $q$ approaches the dust value $1/2$ from below. Its departure from this value reflects the residual contribution of dark energy at finite redshift.

At fixed $\Delta=0.003$, increasing $c$ makes the present acceleration progressively weaker. The present value changes from $q_0=-0.6439$ for $c=0.60$ to $-0.5505$, $-0.4851$, and $-0.3793$ for $c=0.68$, $0.75$, and $0.90$, respectively. The corresponding transition redshifts are $z_t\simeq0.577$, $0.577$, $0.573$, and $0.552$. Thus, the transition epoch remains nearly unchanged between $c=0.60$ and $0.68$ and shifts only moderately thereafter, despite the appreciable variation in the present value of $q$. A substantial change in the current acceleration therefore need not be accompanied by a comparable displacement of its onset.

Panel (b) shows a similar distinction between the present acceleration and the transition epoch when $\Delta$ is varied at fixed $c=0.80$. Standard HDE gives $q_0=-0.5821$ and $z_t\simeq0.578$. For $\Delta=0.001$, $0.002$, and $0.003$, the present values become $q_0=-0.5324$, $-0.4869$, and $-0.4454$, while the corresponding transition redshifts are $z_t\simeq0.576$, $0.573$, and $0.567$. Increasing the deformation therefore weakens the present acceleration and moves its onset to slightly lower redshift for this fixed value of $c$.

The inset shows that the deformation does not generate a uniform vertical displacement of the deceleration curve. Relative to HDE, $q-q^{\rm HDE}$ is positive on the lower redshift side of the transition and becomes negative at higher redshift. This change of sign follows from the dependence of $q$ on the product $w_d\Omega_d$. At positive redshift, increasing $\Delta$ makes $w_d$ less negative while raising $\Omega_d$ relative to HDE. These two changes affect $w_d\Omega_d$ in opposite directions and partly compensate near the acceleration transition. At the present epoch, by contrast, all models share the same value of $\Omega_{d0}$, so the variation of $q_0$ follows directly from the change in $w_{d0}$.

Over the displayed negative redshift interval, $q$ continues to decrease and some representative trajectories reach $q<-1$, which locally corresponds to $\dot H>0$. This behavior is distinct from the onset of acceleration at $q=0$. It describes the finite interval evolution of the NBHDE model when the second-order density is retained, but it does not establish the asymptotic fate of the solutions or the accuracy of the entropy expansion at arbitrarily late times. For the representative cases shown in Fig.~\ref{fig:q_evolution}, the present acceleration varies appreciably, whereas the transition redshift changes more moderately.

\subsection{Background dynamics in the \texorpdfstring{$(c,\Delta)$}{(c, Delta)} parameter space}
\label{subsec:two_parameter}

\begin{figure*}
	\centering
	\includegraphics[width=0.92\textwidth]{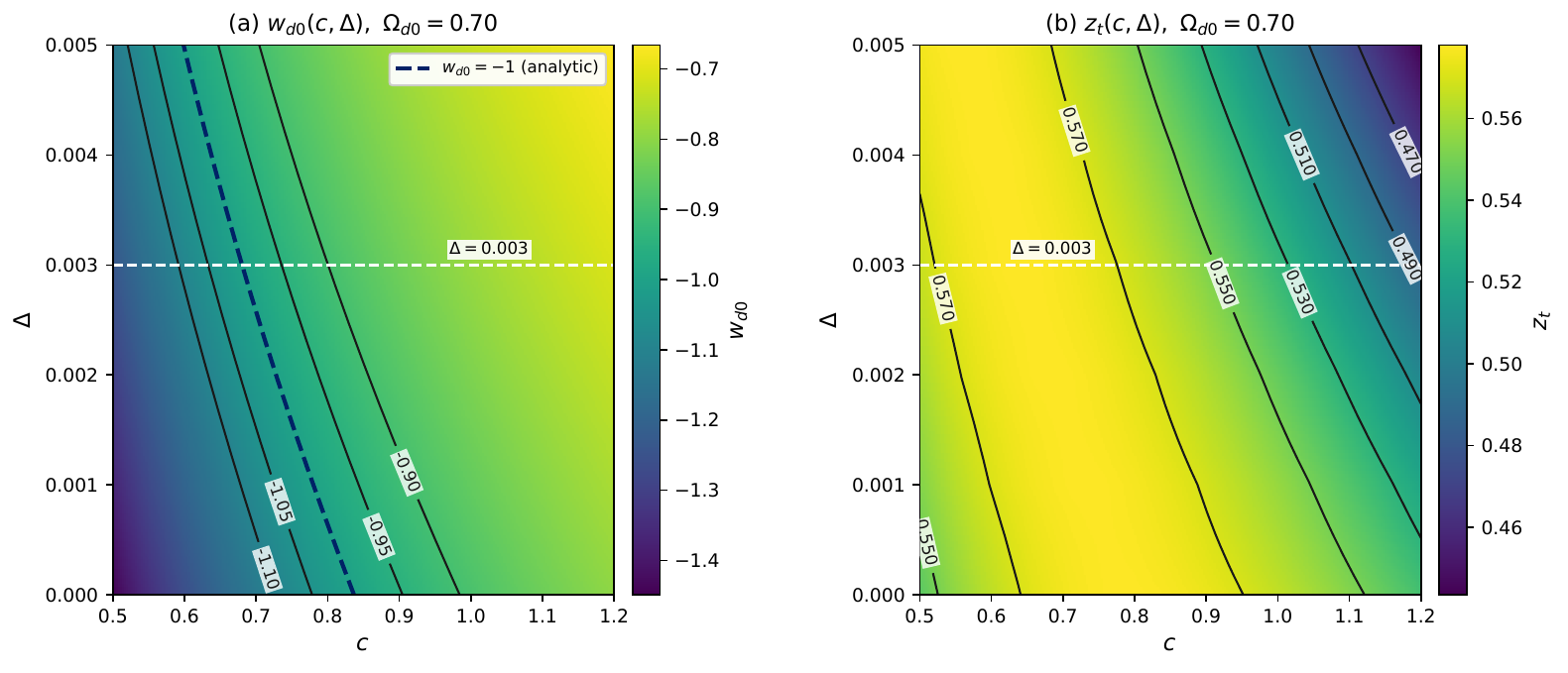}
	\caption{Background quantities across the $(c,\Delta)$ parameter space for $\Omega_{d0}=0.70$. Panel (a) shows the present equation of state $w_{d0}$, and panel (b) shows the acceleration transition redshift $z_t$. The solid black curves are contours of the corresponding quantities. In panel (a), the dark blue dashed curve is the analytic present phantom boundary $w_{d0}=-1$ given by Eq.~\eqref{eq:c_phantom_divide}. The white horizontal dashed line marks $\Delta=0.003$; the region above it is included only as an extended illustration beyond the conservative domain.}
	\label{fig:two_parameter_maps}
\end{figure*}

The one parameter slices considered above do not show how changes in $c$ can compensate for changes in $\Delta$. Figure~\ref{fig:two_parameter_maps} therefore maps the present equation of state and the acceleration transition redshift over $0.50\leq c\leq1.20$ and $0\leq\Delta\leq0.005$, with the background normalization fixed by Eq.~\eqref{eq:present_normalization}. For every parameter pair, the present horizon radius is determined independently from the corresponding NBHDE density constraint.

The horizontal line at $\Delta=0.003$ marks the upper boundary of the conservative domain adopted in Sec.~\ref{subsec:numerical_setup}. The region above this line is retained only to display how the trends continue beyond that domain. The line is not a contour of constant approximation error, and the black contours in both panels represent theoretical values of the plotted quantities rather than observational confidence regions.
Panel (a) shows that increasing either $c$ or $\Delta$ moves $w_{d0}$ toward less negative values. An increase in the deformation can therefore be offset by a decrease in the holographic parameter while approximately preserving the present equation of state. The analytic phantom boundary derived in Eq.~\eqref{eq:c_phantom_divide} follows the numerical separation between the phantom and nonphantom regions. It occurs at $c\simeq0.8367$ in the HDE limit and shifts to $c\simeq0.7799$, $0.7278$, and $0.6804$ for $\Delta=0.001$, $0.002$, and $0.003$, respectively. At fixed $\Delta$, values of $c$ below this boundary give $w_{d0}<-1$, whereas values above it give $w_{d0}>-1$.

This boundary classifies the dark energy only at the present epoch. It does not determine whether the equation of state crosses $w_d=-1$ at another redshift, and parameter pairs with the same $w_{d0}$ need not produce identical functions $w_d(z)$. Moreover, because $\Omega_{d0}$ is fixed, Eq.~\eqref{eq:deceleration_parameter} determines $q_0$ directly from $w_{d0}$. A separate map of the present deceleration parameter would therefore contain no independent information under the adopted normalization.

The dependence of $z_t$ in panel (b) is qualitatively different. At fixed $\Delta$, the transition redshift initially rises with $c$, reaches a broad maximum near $z_t\simeq0.578$, and then decreases. Within the conservative domain, the location of this maximum shifts from approximately $c=0.79$ at $\Delta=0$ to $c=0.64$ at $\Delta=0.003$. The response to $\Delta$ also changes across the plane: increasing the deformation raises $z_t$ near the lower end of the displayed $c$ range but lowers it at larger $c$. The decrease obtained at fixed $c=0.80$ in Sec.~\ref{subsec:q_transition} is consequently not a universal trend over the full parameter plane.

The different contour geometries in the two panels show that a parameter displacement that preserves $w_{d0}$ does not, in general, preserve the acceleration transition redshift. The compensation between $c$ and $\Delta$ is therefore dependent on the cosmological quantity being considered. At this stage it is a property of the theoretical background solutions at fixed $(H_0,\Omega_{m0})$ and should not yet be interpreted as an observational parameter degeneracy.

Table~\ref{tab:representative_nbhde} collects the quantities used in Figs.~\ref{fig:omega_evolution}--\ref{fig:q_evolution}. The first four rows isolate the effect of $\Delta$ at fixed $c=0.80$, while the remaining rows isolate the effect of $c$ at fixed $\Delta=0.003$. Positive and negative values of $z_{\rm ph}$ denote crossings in the past and within the displayed future interval, respectively. A dash indicates that no crossing is found over $-0.99\leq z\leq5$.
\begin{table}[H]
	\centering
	\caption{Representative HDE and NBHDE background quantities for the normalization in Eq.~\eqref{eq:present_normalization}. The local density diagnostic $100\,\delta_{\rm tr}^{\max}$ is evaluated over $-0.99\leq z\leq5$, whereas the independently normalized expansion rate diagnostic $100\,\delta_E^{\max}$ is evaluated over $0\leq z\leq5$. Both diagnostics are expressed as percentages. A dash denotes the absence of a phantom crossing within $-0.99\leq z\leq5$.}
	\label{tab:representative_nbhde}
	\small
	\setlength{\tabcolsep}{7pt}
	\renewcommand{\arraystretch}{1.2}
	\begin{tabular}{cccccccc}
		\hline\hline
		$c$ & $\Delta$ & $w_{d0}$ & $q_0$ & $z_t$ & $z_{\rm ph}$ & $100\,\delta_{\rm tr}^{\max}$ & $100\,\delta_E^{\max}$ \\
		\hline
		0.80 & 0     & -1.0306 & -0.5821 & 0.578 &  0.093 & 0.000 & 0.0000 \\
		0.80 & 0.001 & -0.9833 & -0.5324 & 0.576 & -0.059 & 0.042 & 0.0030 \\
		0.80 & 0.002 & -0.9399 & -0.4869 & 0.573 & -0.254 & 0.303 & 0.0212 \\
		0.80 & 0.003 & -0.9004 & -0.4454 & 0.567 & -0.588 & 0.923 & 0.0631 \\
		\hline
		0.60 & 0.003 & -1.0894 & -0.6439 & 0.577 &  0.238 & 0.917 & 0.0695 \\
		0.68 & 0.003 & -1.0004 & -0.5505 & 0.577 &  0.001 & 0.919 & 0.0668 \\
		0.75 & 0.003 & -0.9382 & -0.4851 & 0.573 & -0.264 & 0.921 & 0.0646 \\
		0.90 & 0.003 & -0.8374 & -0.3793 & 0.552 & --     & 0.933 & 0.0602 \\
		\hline\hline
	\end{tabular}
\end{table}
The table provides a concrete example of the compensation between the two parameters. The combinations $(c,\Delta)=(0.80,0.002)$ and $(0.75,0.003)$ give nearly equal values of $w_{d0}$ and $q_0$, while their transition redshifts agree at the quoted precision. Their phantom crossing redshifts nevertheless differ, confirming that similar present quantities do not make the complete background evolutions identical.

All NBHDE models listed in the table satisfy the adopted subpercent criterion for the local density difference. The largest expansion rate difference among these representative cases is $0.0695\%$ over $0\leq z\leq5$, while the maximum over the full conservative parameter scan is the $0.0732\%$ value reported in Sec.~\ref{subsec:numerical_setup}. As shown in Sec.~\ref{subsec:eos_phantom}, these small background differences do not guarantee identical phantom classifications for models lying very close to the divide.

The parameter maps thus reveal a clear compensation direction between the Barrow deformation and the holographic parameter, but the direction and its effectiveness depend on the background quantity under consideration. Whether the same behavior appears when the expansion history is confronted with data is examined next using CC measurements.

\section{Conditional comparison with CC measurements}
\label{sec:cc_analysis}

The results of Sec.~\ref{cosmo_evol} show that the Barrow deformation can produce appreciable changes in the dark energy equation of state and the phantom crossing epoch while affecting the total expansion rate more weakly. CC measurements provide a direct way to examine how these changes appear in the observed late time expansion history. We therefore compare the NBHDE predictions with a compilation of spectroscopic measurements of $H(z)$ and examine the dependence of the corresponding CC statistic on $c$ and $\Delta$.

Throughout this section, $H_0$, $\Omega_{m0}$, and $\Omega_{d0}$ are held fixed at the values specified in Eq.~\eqref{eq:present_normalization}. The model parameters are varied over $0.50\leq c\leq1.20$ and $0\leq\Delta\leq0.003$, where the upper limit on $\Delta$ is the conservative validity boundary established in Sec.~\ref{subsec:numerical_setup}. All values of $\chi^2_{\rm CC}$ reported below are consequently conditional on this background normalization. The comparison determines how strongly the CC expansion history responds to the two model parameters and whether the compensation identified in Sec.~\ref{subsec:two_parameter} persists at the level of the conditional statistic.

\subsection{CC data and statistics}
\label{subsec:cc_data}

The adopted sample contains 36 spectroscopic CC measurements spanning $0.07\leq z\leq1.965$. These measurements are obtained from differential age estimates and are drawn from Refs.~\cite{Zhang2014,Simon2005,Stern2010,Moresco2012,Moresco2015,Moresco2016,Ratsimbazafy2017,Jiao2023,Tomasetti2023,Loubser2025,Loubser2025DESI}. Estimates of $H(z)$ derived from BAO or other clustering observables are not included. The complete data vector, together with the reported uncertainties and original references, is given in Appendix~\ref{app:cc_data}.

For each pair $(c,\Delta)$, the theoretical prediction is $H_{\rm th}(z)=H_0E(z)$, where $E(z)$ is defined in Eq.~\eqref{eq:num_variables}. When statistical and systematic uncertainties are reported separately and symmetrically, they are combined in quadrature. The measurement of Jiao et al.~\cite{Jiao2023} at $z_{\rm J}=0.8$ is treated separately because its systematic uncertainty is asymmetric. For its central value $H_{\rm J}=113.1\,{\rm km\,s^{-1}\,Mpc^{-1}}$, we use $\sigma_{{\rm J},-}=(15.1^2+11.3^2)^{1/2}$ and $\sigma_{{\rm J},+}=(15.1^2+29.1^2)^{1/2}$. Its contribution to the statistic is
\begin{equation}\label{eq:cc_jiao_chi2}
	\chi^2_{\rm J}=
	\begin{cases}
		\displaystyle\left[\frac{H_{\rm th}(z_{\rm J})-H_{\rm J}}{\sigma_{{\rm J},-}}\right]^2, & H_{\rm th}(z_{\rm J})<H_{\rm J},\\[0.35cm]
		\displaystyle\left[\frac{H_{\rm th}(z_{\rm J})-H_{\rm J}}{\sigma_{{\rm J},+}}\right]^2, & H_{\rm th}(z_{\rm J})\geq H_{\rm J}.
	\end{cases}
\end{equation}

The remaining 35 measurements form the Gaussian data vector. Their covariance matrix $\mathbf{C}_{\rm CC}$ contains the individual variances and the correlated block associated with the measurements at $z=0.46$, $0.67$, and $0.83$ \cite{Loubser2025DESI}. For this ordered redshift triplet, the block is constructed as $(C_{\rm DESI})_{ij}=\rho_{ij}^{\rm pub}\sigma_{i,{\rm tot}}\sigma_{j,{\rm tot}}$, using the published correlation matrix
\begin{equation}\label{eq:cc_desi_correlation}
	\boldsymbol{\rho}_{\rm DESI}^{\rm pub}=
	\begin{pmatrix}
		1 & 0.932 & 0.830 \\
		0.932 & 1 & 0.776 \\
		0.830 & 0.776 & 1
	\end{pmatrix}.
\end{equation}

Writing the residual vector for these 35 measurements as $\Delta\mathbf{H}_{\rm G}=\mathbf{H}_{\rm th,G}-\mathbf{H}_{\rm obs,G}$, the conditional statistic used in the following analysis is
\begin{equation}\label{eq:cc_chi2}
	\chi^2_{\rm CC}=\Delta\mathbf{H}_{\rm G}^{\rm T}\mathbf{C}_{\rm CC}^{-1}\Delta\mathbf{H}_{\rm G}+\chi^2_{\rm J}.
\end{equation}

This construction retains the published asymmetric uncertainty and the available correlation information. Possible correlations between measurements obtained from different surveys, including those associated with common stellar population synthesis modeling, cannot be reconstructed from the published information and are therefore not included \cite{Moresco2020Covariance}. This limitation should be kept in mind when interpreting the absolute value of the conditional statistic.

\subsection{NBHDE expansion histories and CC measurements}
\label{subsec:cc_expansion}

\begin{figure*}
	\centering
	\includegraphics[width=0.92\textwidth]{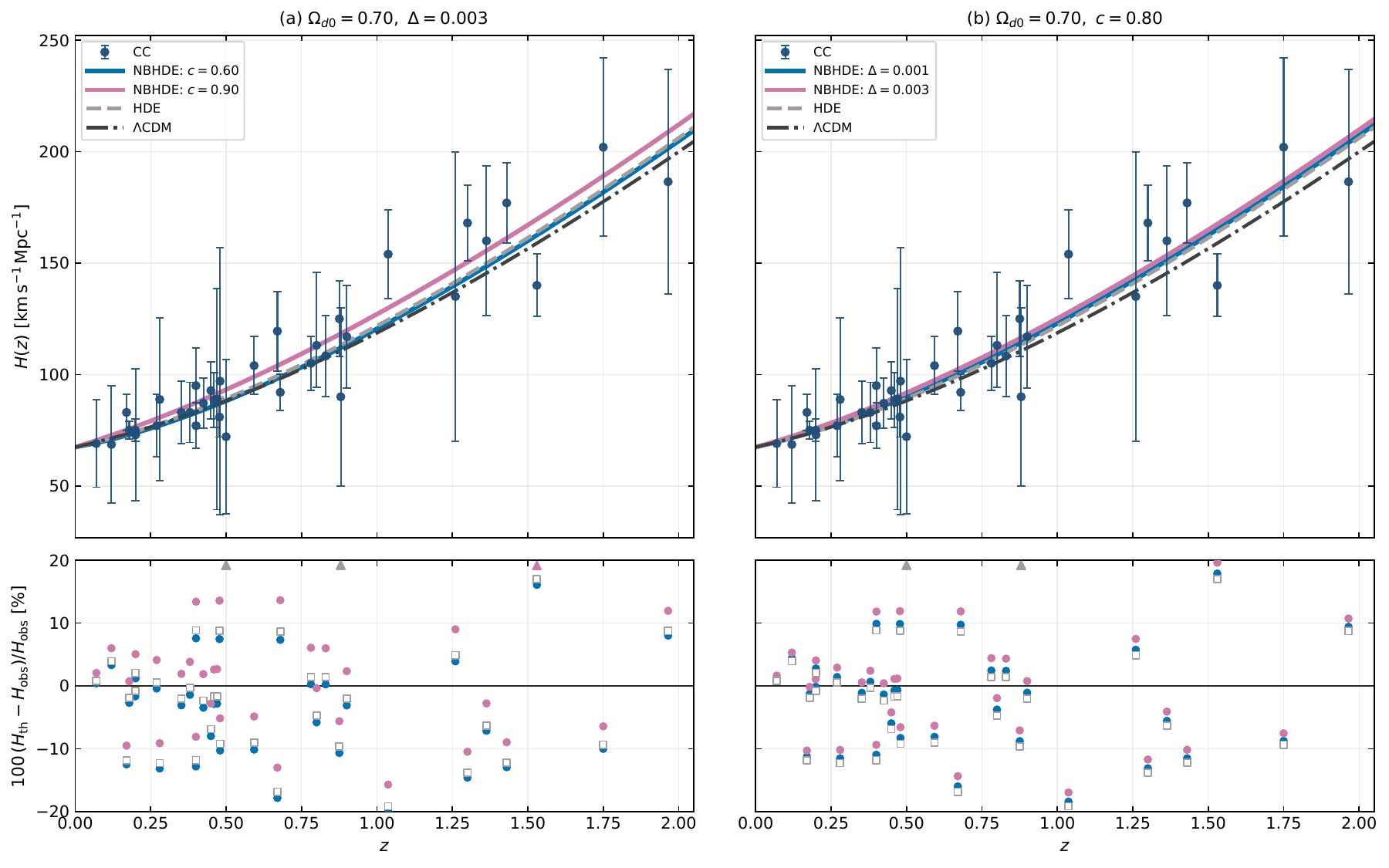}
	\caption{Comparison of representative NBHDE expansion histories with the 36 CC measurements. The upper panels show $H(z)$ together with the reported uncertainties. Panel (a) varies $c$ at fixed $\Delta=0.003$, while panel (b) varies $\Delta$ at fixed $c=0.80$. The HDE curve at $c=0.80$ and a flat $\Lambda$CDM curve with the same fixed background normalization are included for reference. The lower panels show the percentage residuals $100[H_{\rm th}(z_i)-H_i]/H_i$ for the NBHDE and HDE curves. Upward and downward triangles mark residuals outside the displayed range of $\pm20\%$. The NBHDE curves are representative parameter slices and are not selected from the conditional minima discussed in Sec.~\ref{subsec:cc_profile}.}
	\label{fig:ohd_confrontation}
\end{figure*}

Figure~\ref{fig:ohd_confrontation} compares representative NBHDE expansion histories with the CC measurements. The parameter choices are selected to isolate the separate effects of $c$ and $\Delta$. Panel (a) fixes $\Delta=0.003$ and compares $c=0.60$ with $c=0.90$, whereas panel (b) fixes $c=0.80$ and compares $\Delta=0.001$ with $\Delta=0.003$. Both panels also show the HDE prediction at $c=0.80$ and a flat $\Lambda$CDM curve with the same values of $H_0$ and $\Omega_{m0}$. The latter is included only as a familiar reference for the overall shape of the expansion history.
Panel (a) shows that the predicted expansion rate responds appreciably to the holographic parameter. All curves share the same present value $H(0)=H_0$, but their separation increases with redshift. At fixed $\Delta=0.003$, the model with $c=0.90$ predicts a more rapidly rising $H(z)$ than the model with $c=0.60$. The dependence on $c$ is therefore visible not only in the present dark energy equation of state but also in the shape of the expansion history across the CC redshift interval.

The effect of $\Delta$ at fixed $c$ is considerably smaller. In panel (b), the NBHDE curves for $\Delta=0.001$ and $0.003$ remain close to one another and to the HDE curve throughout the observed interval. Increasing $\Delta$ produces a systematic upward shift in $H(z)$ at positive redshift, but the separation is much weaker than that generated by the variation of $c$ shown in panel (a). This contrast is physically relevant because the same change in $\Delta$ produces a much clearer displacement of $w_d(z)$ and of the phantom crossing epoch in Sec.~\ref{subsec:eos_phantom}. The expansion rate and the dark energy equation of state therefore do not respond with equal sensitivity to the Barrow deformation.

The percentage residuals in the lower panels display the same ordering among the theoretical curves. They are defined relative to the measured central values and are included to make the small differences between the models visible. Since these residuals are not normalized by the measurement uncertainties and do not incorporate the correlated covariance block, their magnitudes should not be interpreted as individual contributions to $\chi^2_{\rm CC}$. The relatively large scatter of the CC central values also prevents a reliable ranking of the closely spaced curves by visual inspection alone.

Figure~\ref{fig:ohd_confrontation} thus provides a direct graphical indication of the parameter dependence: over the ranges shown, the predicted expansion history across the CC redshift interval responds more strongly to changes in $c$ than to the conservative variation of $\Delta$. A quantitative assessment of this compensation requires the statistic defined in Eq.~\eqref{eq:cc_chi2}, which is examined across the model parameter space in the following subsection.

\subsection{CC statistic in the model parameter space}
\label{subsec:cc_profile}

We now examine the conditional statistic over the model parameter space specified above. At each fixed value of $\Delta$, we determine the value of the holographic parameter that minimizes $\chi^2_{\rm CC}$:

\begin{equation}\label{eq:cc_minimum_definition}
	c_{\min}(\Delta)=\operatorname*{arg\,min}_{0.50\leq c\leq1.20}\chi^2_{\rm CC}(c,\Delta).
\end{equation}

\begin{figure}
	\centering
	\includegraphics[width=0.5\columnwidth]{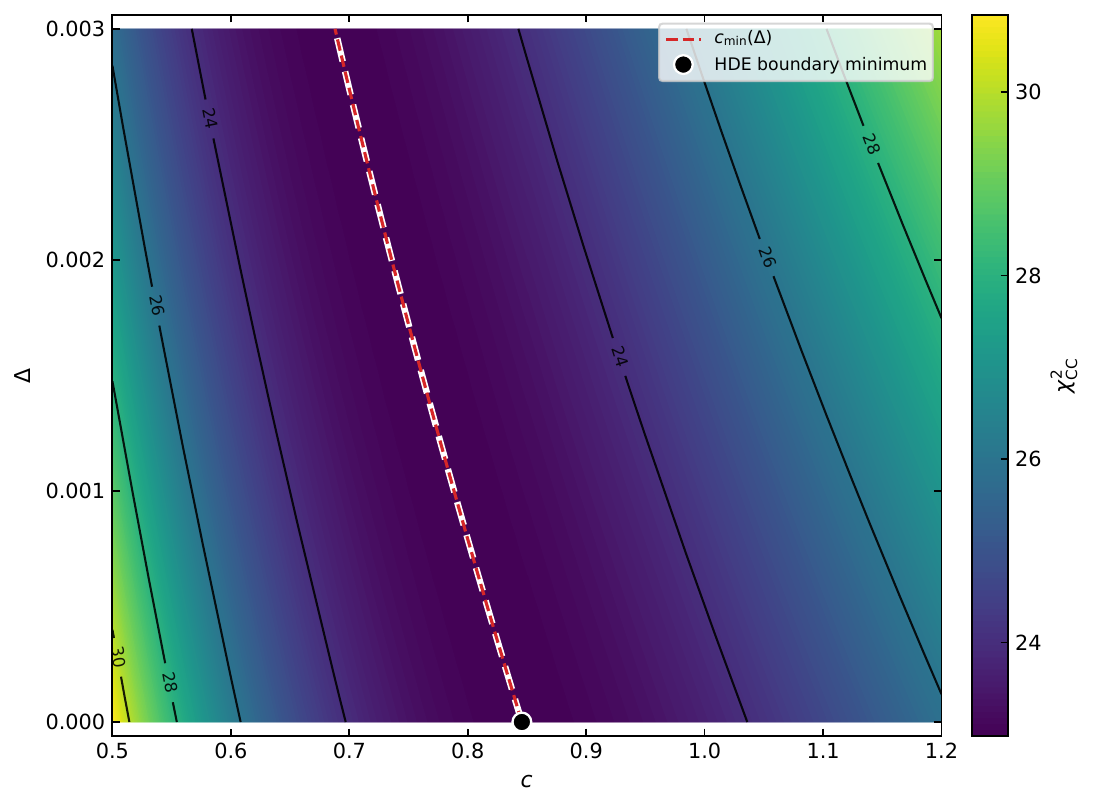}
	\caption{Conditional $\chi^2_{\rm CC}(c,\Delta)$ surface for the fixed background normalization in Eq.~\eqref{eq:present_normalization}. The red dashed curve shows $c_{\min}(\Delta)$, the value of $c$ that minimizes the statistic at each fixed $\Delta$, and the black circle marks the overall conditional minimum on the HDE boundary. The black contours represent raw values of $\chi^2_{\rm CC}$ and are not confidence contours.}
	\label{fig:cc_chi2_map}
\end{figure}

The resulting $\chi^2_{\rm CC}$ surface and the curve of conditional minima are shown in Fig.~\ref{fig:cc_chi2_map}.
The surface contains an elongated region of low $\chi^2_{\rm CC}$ extending diagonally across the full deformation interval. Its lowest point occurs at $c=0.845686$ and $\Delta=0$, where $\chi^2_{\rm CC,min}=22.983589$. Since NBHDE reduces exactly to HDE at $\Delta=0$, the marked point is a single boundary minimum common to both descriptions. Its recovery also provides a numerical check of the HDE limit established in Sec.~\ref{subsec:nbhde_hde_limit}.

As $\Delta$ increases, the conditional minimum shifts continuously toward smaller values of $c$. Table~\ref{tab:cc_profile} reports representative points along this curve. The quantity $\Delta\chi^2_{\rm CC}$ is defined relative to the overall boundary minimum as $\chi^2_{\rm CC}[c_{\min}(\Delta),\Delta]-\chi^2_{\rm CC,min}$.
\begin{table}[H]
	\centering
	\caption{Conditional CC minima at fixed $\Delta$ for the background normalization in Eq.~\eqref{eq:present_normalization}. The last column gives the change relative to the overall minimum at $\Delta=0$.}
	\label{tab:cc_profile}
	\small
	\setlength{\tabcolsep}{7.0pt}
	\renewcommand{\arraystretch}{1.2}
	\begin{tabular}{cccc}
		\hline\hline
		$\Delta$ & $c_{\min}$ & $\chi^2_{\rm CC}$ & $\Delta\chi^2_{\rm CC}$ \\
		\hline
		0.0000 & 0.845686 & 22.983589 & 0 \\
		0.0010 & 0.788401 & 22.983944 & $3.55\times10^{-4}$ \\
		0.0020 & 0.735798 & 22.984284 & $6.96\times10^{-4}$ \\
		0.0030 & 0.687940 & 22.984601 & $1.01\times10^{-3}$ \\
		\hline\hline
	\end{tabular}
\end{table}
Across the conservative deformation interval, $c_{\min}$ decreases from $0.8457$ to $0.6879$, whereas the conditional statistic changes by only about $10^{-3}$. A reduction in $c$ therefore offsets most of the change in the expansion history produced by increasing $\Delta$. The compensation seen in the theoretical background quantities in Sec.~\ref{subsec:two_parameter} consequently remains present when the model is compared with the CC expansion history.

We examined two alternative treatments of the three correlated measurements to assess whether this behavior is driven by the adopted covariance block. Replacing that block by its diagonal part or removing the three measurements changes both the absolute value of $\chi^2_{\rm CC}$ and the corresponding values of $c_{\min}$. In either case, however, the change in the minimum statistic between $\Delta=0$ and $0.003$ remains smaller than $10^{-4}$ in magnitude. These limited checks show that the compensation direction is not produced solely by the treatment of this three point covariance block, although the principal results in Fig.~\ref{fig:cc_chi2_map} and Table~\ref{tab:cc_profile} retain the published correlations.

The effect of the second-order truncation on the conditional comparison is also small. At $\Delta=0.003$, repeating the calculation with the exact BHDE density and minimizing independently over $c$ gives $c_{\min}^{\rm ex}=0.684780$, with a minimum statistic of $\chi^2_{\rm CC}=22.984665$. This value exceeds the corresponding NBHDE minimum by only $6.36\times10^{-5}$. Thus, at the upper boundary of the adopted deformation range, the residual density truncation has a negligible numerical effect on the conditional CC statistic.

Within this conditional comparison, a nonzero deformation does not lower $\chi^2_{\rm CC}$ below its value on the HDE boundary. At the same time, the increase in the statistic along the curve of conditional minima is extremely small. The CC expansion history therefore provides little separation between HDE and nearby NBHDE models once the compensating change in $c$ is taken into account. This result does not constitute an observational bound on $\Delta$; rather, it shows that differences in $w_d(z)$ and in the phantom crossing epoch can remain appreciable even when the corresponding expansion histories give nearly identical conditional CC statistics.

\section{Conclusions and outlook}
\label{conclu}

In this work, we formulated NBHDE as a controlled second-order realization of the Barrow entropy correction near the standard HDE limit. By normalizing the correction through the entropy ratio $S_B/S_{\rm BH}$ and expanding the Barrow power law through second-order, the model acquires logarithmic terms whose coefficients are fixed by the underlying entropy rather than introduced phenomenologically. With the future event horizon as the infrared cutoff, NBHDE provides a continuous extension of HDE in a spatially flat universe containing pressureless matter and dark energy, without an independent cosmological constant.

The formulation is mathematically self-consistent at the background level. Its energy density is strictly positive, the present density constraint admits a unique positive horizon radius, and the standard HDE equations are recovered exactly at $\Delta=0$. The local horizon equation also satisfies the terminal condition required by the integral definition of the future event horizon when the second-order density is retained throughout the future evolution. Independently of these consistency properties, we quantified the accuracy of the entropy expansion by comparing NBHDE with an exact density reference calculation normalized through its own present density constraint. Over the scanned range $0.50\leq c\leq1.20$, the interval $0\leq\Delta\leq0.003$ defines a conservative domain in which the maximum local density difference remains below $1\%$ over $-0.99\leq z\leq5$, while the maximum relative difference in the normalized Hubble rate remains below $0.074\%$ over $0\leq z\leq5$. The negative redshift results outside this validated comparison interval describe the evolution of NBHDE itself and are not used to extend the accuracy claim to arbitrarily late times.

Within this domain, the model evolves regularly from matter domination to dark energy domination and accelerated expansion. The Barrow correction produces comparatively small changes in the total expansion rate and the acceleration transition, but it has a more pronounced effect on the dark energy equation of state and the phantom crossing. At $c=0.80$, for example, the present equation of state changes from $w_{d0}=-1.0306$ in the HDE limit to $w_{d0}=-0.9004$ at $\Delta=0.003$, moving the present solution from the phantom to the nonphantom regime. Over the same interval, the crossing redshift shifts from $z_{\rm ph}\simeq0.093$ to $z_{\rm ph}\simeq-0.588$. The results summarized in Fig.~\ref{fig:w_evolution} and Table~\ref{tab:representative_nbhde} therefore show that a deformation producing only a small change in $H(z)$ can nevertheless alter the physical classification and temporal evolution of the dark energy component.

The two dimensional analysis in Fig.~\ref{fig:two_parameter_maps} further reveals a correlated but observable dependent response to $c$ and $\Delta$. Increasing either parameter moves the present equation of state toward less negative values, so part of the effect of increasing $\Delta$ can be offset by decreasing $c$. The analytic present phantom boundary agrees with the numerical separation between the phantom and nonphantom regions. However, the contour geometries show that a displacement preserving $w_{d0}$ does not generally preserve the acceleration transition redshift. Likewise, the representative models in Table~\ref{tab:representative_nbhde} demonstrate that nearly equal present values of $w_d$ and $q$ do not require identical phantom crossing redshifts. The compensation between $c$ and $\Delta$ is therefore a property of particular cosmological quantities rather than a universal equivalence between the corresponding background solutions.

The conditional comparison with 36 spectroscopic CC measurements provides an observational counterpart to this dynamical behavior. At fixed $H_0=67.4\,{\rm km\,s^{-1}\,Mpc^{-1}}$ and $\Omega_{m0}=0.30$, the lowest value of the conditional statistic occurs on the HDE boundary at $c=0.845686$ and $\Delta=0$, where $\chi^2_{\rm CC,min}=22.983589$. At $\Delta=0.003$, the minimum shifts to $c_{\min}=0.687940$, while the statistic increases by only $\Delta\chi^2_{\rm CC}=1.01\times10^{-3}$. Figure~\ref{fig:cc_chi2_map} and Table~\ref{tab:cc_profile} thus quantify how a decrease in $c$ offsets the effect of the Barrow deformation on the observed expansion history. Within this fixed normalization analysis, the CC data show no preference for a nonzero deformation, but they also provide very little separation between the HDE boundary and nearby NBHDE models once $c$ is adjusted. Repeating the calculation at $\Delta=0.003$ with the exact BHDE density changes the corresponding minimum statistic by only $6.36\times10^{-5}$, confirming that this conditional result is not driven by the residual second-order truncation error.

Taken together, these results establish NBHDE as a well defined and quantitatively controlled framework for studying the leading Barrow corrections near the HDE limit. Its value lies in separating three logically distinct questions: the internal consistency of the second-order model, its numerical accuracy relative to the unexpanded Barrow density, and its physical departure from standard HDE. The exact density reference calculation supplies a quantitative measure of the truncation error, while the second-order formulation makes the logarithmic structure and its dynamical consequences directly accessible. A natural next step is a joint background analysis in which $H_0$, $\Omega_{m0}$, $c$, and $\Delta$ are varied simultaneously using complementary CC, Type Ia supernova, and BAO measurements. A subsequent formulation of linear perturbations would permit the inclusion of the full CMB information, redshift space distortions, and weak lensing observables. These extensions can test whether the combination of geometric and growth measurements breaks the compensation between $c$ and $\Delta$ identified here, building on the controlled background foundation established in the present work.

\appendix
\section{Spectroscopic CC data compilation}
\label{app:cc_data}

This appendix lists the 36 spectroscopic CC measurements used in the observational analysis of Sec.~\ref{sec:cc_analysis}. The sample spans the redshift interval $0.07\leq z\leq1.965$ and is restricted to determinations of $H(z)$ obtained from the differential age method. Measurements inferred from BAO or other clustering observables are not included.

The compilation is drawn from Refs.~\cite{Zhang2014,Simon2005,Stern2010,Moresco2012,Moresco2015,Moresco2016,Ratsimbazafy2017,Jiao2023,Tomasetti2023,Loubser2025,Loubser2025DESI}. For the measurements reported by Moresco et al.~\cite{Moresco2012,Moresco2016}, we adopt the values calibrated with the BC03 stellar population synthesis model and the corresponding uncertainties given in the original analyses. When statistical and systematic uncertainties are reported separately, both components are retained in Table~\ref{tab:cc36_data} in their published form.

The uncertainty and covariance treatment follows Sec.~\ref{subsec:cc_data}. Symmetric statistical and systematic uncertainties are combined in quadrature when constructing the conditional statistic. For the measurement of Jiao et al.~\cite{Jiao2023} at $z=0.8$, the lower and upper total uncertainties are retained separately through the piecewise contribution in Eq.~\eqref{eq:cc_jiao_chi2}. The measurements at $z=0.46$, $0.67$, and $0.83$ enter the analysis through a correlated covariance block constructed from their total uncertainties and the published correlation coefficients in Eq.~\eqref{eq:cc_desi_correlation}. Correlations between measurements from different surveys cannot be included because a complete public cross survey covariance matrix is not available \cite{Moresco2020Covariance}.
\begin{table}[H]
	\centering
	\caption{Spectroscopic CC measurements adopted in this work. Values of $H(z)$ are given in ${\rm km\,s^{-1}\,Mpc^{-1}}$. Statistical and systematic uncertainties are listed separately when both components are reported in the original source.}
	\label{tab:cc36_data}
	\small
	\setlength{\tabcolsep}{3.0pt}
	\renewcommand{\arraystretch}{1.1}
	\begin{tabular*}{0.7\textwidth}{@{\extracolsep{\fill}}lll@{\hspace{0.7cm}}lll@{}}
		\hline\hline
		$z$ & $H(z)$ & Refs. & $z$ & $H(z)$ & Refs. \\
		\hline
		0.0700 & $69.0\pm19.6$ & \cite{Zhang2014} & 0.5000 & $72.1\pm33.9_{\rm stat}\pm7.3_{\rm syst}$ & \cite{Loubser2025} \\
		0.1200 & $68.6\pm26.2$ & \cite{Zhang2014} & 0.5929 & $104\pm13$ & \cite{Moresco2012} \\
		0.1700 & $83\pm8$ & \cite{Simon2005} & 0.6700 & $119.45\pm6.39_{\rm stat}\pm16.64_{\rm syst}$ & \cite{Loubser2025DESI} \\
		0.1791 & $75\pm4$ & \cite{Moresco2012} & 0.6797 & $92\pm8$ & \cite{Moresco2012} \\
		0.1993 & $75\pm5$ & \cite{Moresco2012} & 0.7812 & $105\pm12$ & \cite{Moresco2012} \\
		0.2000 & $72.9\pm29.6$ & \cite{Zhang2014} & 0.8000 & $113.1\pm15.1_{\rm stat}\,{}^{+29.1}_{-11.3}{}_{\rm syst}$ & \cite{Jiao2023} \\
		0.2700 & $77\pm14$ & \cite{Simon2005} & 0.8300 & $108.28\pm10.07_{\rm stat}\pm15.08_{\rm syst}$ & \cite{Loubser2025DESI} \\
		0.2800 & $88.8\pm36.6$ & \cite{Zhang2014} & 0.8754 & $125\pm17$ & \cite{Moresco2012} \\
		0.3519 & $83\pm14$ & \cite{Moresco2012} & 0.8800 & $90\pm40$ & \cite{Stern2010} \\
		0.3802 & $83.0\pm13.5$ & \cite{Moresco2016} & 0.9000 & $117\pm23$ & \cite{Simon2005} \\
		0.4000 & $95\pm17$ & \cite{Simon2005} & 1.0370 & $154\pm20$ & \cite{Moresco2012} \\
		0.4004 & $77.0\pm10.2$ & \cite{Moresco2016} & 1.2600 & $135\pm65$ & \cite{Tomasetti2023} \\
		0.4247 & $87.1\pm11.2$ & \cite{Moresco2016} & 1.3000 & $168\pm17$ & \cite{Simon2005} \\
		0.4497 & $92.8\pm12.9$ & \cite{Moresco2016} & 1.3630 & $160\pm33.6$ & \cite{Moresco2015} \\
		0.4600 & $88.48\pm0.57_{\rm stat}\pm12.32_{\rm syst}$ & \cite{Loubser2025DESI} & 1.4300 & $177\pm18$ & \cite{Simon2005} \\
		0.4700 & $89\pm23_{\rm stat}\pm44_{\rm syst}$ & \cite{Ratsimbazafy2017} & 1.5300 & $140\pm14$ & \cite{Simon2005} \\
		0.4783 & $80.9\pm9.0$ & \cite{Moresco2016} & 1.7500 & $202\pm40$ & \cite{Simon2005} \\
		0.4800 & $97\pm60$ & \cite{Stern2010} & 1.9650 & $186.5\pm50.4$ & \cite{Moresco2015} \\
		\hline
	\end{tabular*}
\end{table}
The measurements in Table~\ref{tab:cc36_data} form the complete observational vector used in Fig.~\ref{fig:ohd_confrontation} and in the conditional statistic of Eq.~\eqref{eq:cc_chi2}. Their comparison with the NBHDE expansion histories is discussed in Sec.~\ref{subsec:cc_expansion}, while the behavior of the conditional minimum across the model parameter space is presented in Sec.~\ref{subsec:cc_profile}. The interpretation of this compilation is therefore restricted to the fixed normalization and parameter domain specified in Sec.~\ref{sec:cc_analysis}.


\clearpage
\bibliography{Ref_NBHDE}
\bibliographystyle{unsrt}

\end{document}